\documentclass{ws-ijmpe}

\begin{document}

\markboth{Xuguang Huang, Xuewen Hao and Pengfei Zhuang}{Asymmetric
Fermi Superfluid With Two Types Of Pairings}

\catchline{}{}{}{}{}

\title{Asymmetric Fermi Superfluid With Two Types Of Pairings}

\author{Xuguang Huang, Xuewen Hao and Pengfei Zhuang}

\address{Physics Department, Tsinghua University\\
Beijing, 100084,
P.R.China }

\maketitle


\begin{abstract}
We investigate the phase diagram in the plane of temperature and
chemical potential mismatch for an asymmetric fermion superfluid
with double- and single-species pairings. There is no mixing of
these two types of pairings at fixed chemical potential, but the
introduction of the single species pairing cures the magnetic
instability at low temperature.
\end{abstract}

\section{Introduction}
\ \ \ \ Inspired by ultracold atomic physics, nuclear physics and
color superconductivity, the fermion pairing between different
species with mismatched Fermi surfaces prompted great
interest\cite{zwie,part,shin,son,calson,he1,iskin,bulgac,ho,giannakis,hu,machida1,caldas}
in recent experimental and theoretical studies. In conventional
fermion superfluid the ground state is well described by the BCS
theory, while for asymmetric fermion superfluid with
double-species pairing the phase structure is much more rich and
the pairing mechanism is not yet very clear. Various exotic phases
have been suggested in the literatures, such as the Sarma
phase\cite{sarma,shov,liu}, the
Fulde-Ferrel-Larkin-Ovchinnikov(FFLO) phase\cite{fuld}, the phase
with deformed Fermi surfaces\cite{muth,sedr}, and the phase
separation\cite{beda,cald} in coordinate space.

For asymmetric fermion superfluid, one of the most surprising
phenomena is the magnetic instability, $i.e.$, when the asymmetry
characterized by the chemical potential mismatch is larger than
the condensate of Cooper pair, the BCS superfluid is unstable
under the perturbation of an external magnetic field. This
instability leads to the ground state to be a FFLO phase. For
color superconductivity, a similar but more complicated phenomenon
is also found, namely the chromomagnetic instability\cite{shov}.
One of the mechanisms to cure the magnetic instability is to
introduce another pairing channel. In this paper, we propose a
simple model including both double- and single-species pairings
and investigate how the single-species pairing influences the FFLO
state.

Our model can be used to describe a general fermion superfluid
system where fermions from the same species can form Cooper pairs.
In ultracold atom gas like $^6$Li and $^{40}$K, there exist
pairings between different elements and between different states
of the same element\cite{huang}. In color superconductivity phase
of dense quark matter\cite{huang1}, the quarks of different
flavors can form total spin zero Cooper pairs and the quarks of
the same flavor can combine into total spin one pairs too. In
neutron stars\cite{lomb}, proton-proton, neutron-neutron and
neutron-proton pairs are all possible. Recently discovered high
temperature superconductivity of
MgB$_2$\cite{iava,bouquet,tsuda,geek} can also be well described
by an extended two-band BCS theory where the electrons from the
same energy bands form Cooper pairs.

\section{The Model}
\ \ \ \ We consider a fermion system containing two species $a$
and $b$ with masses $m_a$ and $m_b$ and chemical potentials
$\mu_a$ and $\mu_b$ satisfying $\mu_a<\mu_b$. The system can be
described by the Hamiltonian
\begin{eqnarray}
\label{h1}
\hat{\cal H}&=&\sum_{\bf p,\sigma}\left[\epsilon_a \hat
a_{\bf p}^{\sigma+} \hat a_{\bf p}^\sigma+\epsilon_b \hat b_{\bf
p}^{\sigma+} \hat b_{\bf p}^\sigma\right] \nonumber\\
&&-\sum_{{\bf p},{\bf k}}\left[\frac{g}{2}\sum_{\sigma\sigma'}\hat
a_{\bf p}^{\sigma+}\hat b_{-{\bf p}}^{-\sigma+}\hat b_{-{\bf
k}}^{-\sigma'}\hat a_{\bf k}^{\sigma'}+g_a\hat a_{\bf
p}^{\uparrow+}\hat a_{-{\bf p}}^{\downarrow+}\hat a_{-{\bf
k}}^\downarrow\hat a_{\bf k}^\uparrow +g_b\hat b_{\bf
p}^{\uparrow+}\hat b_{-{\bf p}}^{\downarrow+}\hat b_{-{\bf
k}}^\downarrow\hat b_{\bf k}^\uparrow\right],
\end{eqnarray}
where ${\bf p}$ and ${\bf k}$ are fermion momenta, $\hat a, \hat
b, \hat a^+$ and $\hat b^+$ are the annihilation and creation
operators, the coupling constants $g, g_a$ and $g_b$ are positive
to keep the interactions attractive, $\epsilon_a={\bf
p}^2/(2m_a)-\mu_a$ and $\epsilon_b={\bf p}^2/(2m_b)-\mu_b$ are the
non-relativistic dispersion relations, and in continuous limit we
simply replace $\sum_{\bf p}$ with $\int d^3{\bf p}/(2\pi)^3$. The
(pseudo-)spin $\sigma,\sigma'=\uparrow,\downarrow$ has been
introduced to keep our model satisfying the Pauli rule explicitly.
Other inner structures like isospin, flavor and color are
neglected since they are not essential for our purpose. The
Hamiltonian has the symmetry $U_a(1)\otimes U_b(1)$ with the
element $U(\theta_a,\theta_b)$ defined as
$U(\theta_a,\theta_b)\hat a_{\bf
p}U^+(\theta_a,\theta_b)=e^{i\theta_a}\hat a_{\bf p}$ and $
U(\theta_a,\theta_b)\hat b_{\bf
p}U^+(\theta_a,\theta_b)=e^{i\theta_b}\hat b_{\bf p}$.

In the framework of mean field approximation, after taking a
Bogoliubov-Valatin transformation from fermions $a$ and $b$ to
quasi-particles $A$ and $B$, the thermodynamic potential can be
expressed in terms of the quasi-particles,
\begin{eqnarray}
\label{omega}
\Omega &=& \Delta_a^2/g_a+\Delta_b^2/g_b+2\Delta^2/g\nonumber\\
&&+\sum_{\bf
p}\left[\epsilon_a+\epsilon_b-E_A-E_B-2T\ln\left(\left((1+e^{-E_A/T}\right)\left(1+e^{-E_B/T}\right)\right)\right],
\end{eqnarray}
where \begin{equation} \Delta_a = g_a\sum_{\bf p}\langle \hat
a_{-{\bf p}}^\downarrow\hat a_{\bf p}^\uparrow\rangle,\ \ \Delta_b =
g_b\sum_{\bf p}\langle \hat b_{-{\bf p}}^\downarrow\hat b_{\bf
p}^\uparrow\rangle,\ \ \ \Delta = g/2\sum_{\bf p,\sigma}\langle \hat
b_{-{\bf p}}^{-\sigma}\hat a_{\bf p}^\sigma\rangle
\end{equation}
are the corresponding $a$-$a$, $b$-$b$ and $a$-$b$ pairing
condensates, and
\begin{equation}
E_A^2 =
\epsilon_+^2+\sqrt{\left(\epsilon_-^2\right)^2+\left(\epsilon_\Delta^2\right)^2},\
\ \ E_B^2 =
\epsilon_+^2-\sqrt{\left(\epsilon_-^2\right)^2+\left(\epsilon_\Delta^2\right)^2}
\end{equation}
are the quasi-particle energies with
\begin{eqnarray}
\epsilon_\pm^2
&=&\left[\left(\epsilon_a^2+\Delta_a^2+\Delta^2\right)\pm\left(\epsilon_b^2+\Delta_b^2+\Delta^2\right)\right]/2,\nonumber\\
\left(\epsilon_\Delta^2\right)^2&=&\Delta^2\left[\left(\epsilon_a-\epsilon_b\right)^2+\left(\Delta_a-\Delta_b\right)^2\right].
\end{eqnarray}
$\Delta_a\neq 0$ and $\Delta_b\neq 0$ correspond, respectively, to
the spontaneous symmetry breaking patterns $U_a(1)\otimes
U_b(1)\rightarrow U_b(1)$ and $U_a(1)$, and $\Delta\neq 0$ means
the breaking pattern $U_a(1)\otimes U_b(1)\rightarrow U_{a-b}(1)$
with the element $U(\theta)$ defined as $U(\theta)\hat a_{\bf
p}U^+(\theta)=e^{i\theta}\hat a_{\bf p}$ and $ U(\theta)\hat
b_{\bf p}U^+(\theta)=e^{-i\theta}\hat b_{\bf p}$. For simplicity,
we have considered all the condensates as real numbers.

To study the FFLO state, we modify the definitions of the fermion
dispersions as
\begin{equation}
\epsilon_i^\pm=({\bf p}\pm{\bf
q}_i)^2/(2m_i)-\mu_i,\ \ \ i=a,b
\end{equation}
with $+$ corresponding to spin-up and $-$ to spin-down, where
${\bf q}_i$ are the FFLO momenta of condensates of $i-i$ pairing,
which together with the condensates are self-consistently
determined by the coupled set of gap equations
\begin{equation}
\label{gap1}
\partial\Omega/\partial\Delta_a=0,\ \ \
\partial\Omega/\partial\Delta_b=0,\ \ \
\partial\Omega/\partial\Delta=0,\ \ \
\partial\Omega/\partial{\bf q}_i=0
\end{equation}
and the minimum thermodynamic potential. Note that the FFLO state
we considered here is the simplest form where the Cooper pairs in
coordinate space have the plane wave forms
\begin{equation}
\Delta({\bf x})=\Delta e^{i({\bf q}_a+{\bf q}_b)\cdot{\bf x}},\ \
\ \Delta_a({\bf x})=\Delta_a e^{2i{\bf q}_a\cdot{\bf x}},\ \ \
\Delta_b({\bf x})=\Delta_b e^{2i{\bf q}_b\cdot{\bf x}}.
\end{equation}
We will choose ${\bf q}_a={\bf q}_b={\bf q}$, such a choice can
avoid possible dynamic instability due to the different superflow
velocities of $a-a$ and $b-b$ channels\cite{khal,mine,nepo}.
Obviously, the translational symmetry and rotational symmetry in
the FFLO state are spontaneously broken. To have a simple
analytical expression for the spectral of quasi-particles, we set
$m_a=m_b=m$. In this case, the thermodynamic potential reads,
\begin{eqnarray}
\label{omega1} \Omega &=&
\Delta_a^2/g_a+\Delta_b^2/g_b+2\Delta^2/g+\sum_{\bf
p}\Big[\epsilon_a^-+\epsilon_b^--\left(E_A^++E_A^-+E_B^++E_B^-\right)/2\nonumber\\
&&-T\ln\left(\left(1+e^{-E_A^+/T}\right)\left(1+e^{-E_A^-/T}\right)\left(1+e^{-E_B^+/T}\right)\left(1+e^{-E_B^-/T}\right)\right)\Big],
\end{eqnarray}
where
\begin{eqnarray}
E_A^\pm
&=&\sqrt{\epsilon_S^2+\delta\epsilon^2+\sqrt{(\epsilon_A^2)^2+(\epsilon_\Delta^2)^2}}\pm\delta\epsilon,\nonumber\\
E_B^\pm
&=&\sqrt{\epsilon_S^2+\delta\epsilon^2-\sqrt{(\epsilon_A^2)^2+(\epsilon_\Delta^2)^2}}\pm\delta\epsilon
\end{eqnarray}
are the quasi-particle energies with the notations
\begin{eqnarray}
\epsilon_S^2
&=&\left[\left(\epsilon_a^+\epsilon_a^-+\Delta_a^2+\Delta^2\right)+\left(\epsilon_b^+\epsilon_b^-+\Delta_b^2+\Delta^2\right)\right]/2,\nonumber\\
\epsilon_A^2
&=&\left[\left(\epsilon_a^+\epsilon_a^-+\Delta_a^2+\Delta^2\right)-\left(\epsilon_b^+\epsilon_b^-+\Delta_b^2+\Delta^2\right)\right]/2,\nonumber\\
\delta\epsilon&=&(\epsilon_a^+-\epsilon_a^-)/2=(\epsilon_b^+-\epsilon_b^-)/2,\nonumber\\
\left(\epsilon_\Delta^2\right)^2&=&\Delta^2\left[(\epsilon_a^+-\epsilon_b^+)(\epsilon_a^--\epsilon_b^-)+(\Delta_b-\Delta_a)^2\right].
\end{eqnarray}

The magnetic stability, namely the stability of the homogeneous
superfluid against the pair momentum fluctuations is characterized
by the superfluid density
\begin{equation}
\rho_s = \partial^2\Omega/
\partial{\bf q}^2\big|_{{\bf q}=0}.
\end{equation}
Positive $\rho_s$ means stable homogeneous superfluid and negative
$\rho_s$ indicates possible FFLO state. In this sense $\bf q$
plays the role of an external magnetic potential $e{\bf A}$.

In the end of this section, we calculate the gapless nodes for the
case with zero FFLO momentum where at least one of $E_A^\pm$ or
$E_B^\pm$ crosses the momentum axis,
\begin{eqnarray}
\label{neq}
E_A^+E_A^-
E_B^+E_B^-&=&\left(\Delta_a\epsilon_b-\Delta_b\epsilon_a\right)^2+\left(\Delta^2+\Delta_a\Delta_b+\epsilon_a\epsilon_b\right)^2=0.
\end{eqnarray}
There are two classes of gapless solutions. 1) Only $\Delta$ is
nonzero. The gapless mode happens at momenta
\begin{equation}
p_\pm=\sqrt{2m\left(\mu\pm\sqrt{\delta\mu^2-\Delta^2}\right)},
\end{equation}
where we have introduced the average chemical potential
$\mu=\left(\mu_a+\mu_b\right)/2$ and the chemical potential
mismatch $\delta\mu=\left(\mu_b-\mu_a\right)/2$. It is clear that
the condition to have the gapless mode is $\delta\mu>\Delta$. 2)
All the three condensates are zero or only one of $\Delta_a$ and
$\Delta_b$ is nonzero. In these cases, we have $\epsilon_a=0$ or
$\epsilon_b=0$ or $\epsilon_a\epsilon_b=0$ which mean real fermion
excitations exactly at the Fermi surfaces.

\section{Phase Diagrams}
\ \ \ \ Since the model is non-renormalizable, we introduce in the
numerical calculations a cutoff $\Lambda=\lambda p_F$ with
$\lambda=2^{1/4}$ and the average Fermi momentum $p_F=\sqrt{2m\mu}$.
We choose $p_Fa=0.4,\ p_Fa_a=p_Fa_b=0.32$ with $a=mg/(4\pi),
a_a=mg_a/(4\pi)$ and $a_b=mg_b/(4\pi)$ being the $s$-wave scattering
lengths. We have checked that in the parameter region of $0<p_Fa,\
p_Fa_a,\ p_Fa_b <1$ and $0<a_a=a_b<a$ there is no qualitative change
in the phase diagrams.

\begin{figure}[!htb]
\begin{center}
\includegraphics[width=8cm]{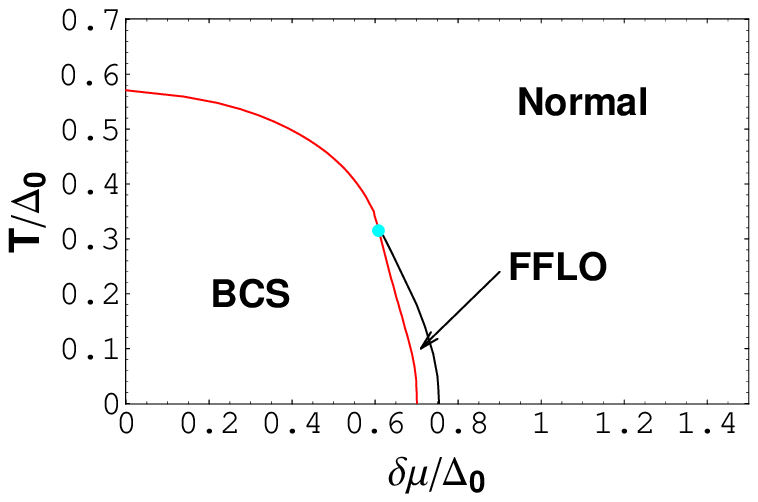}
\includegraphics[width=8cm]{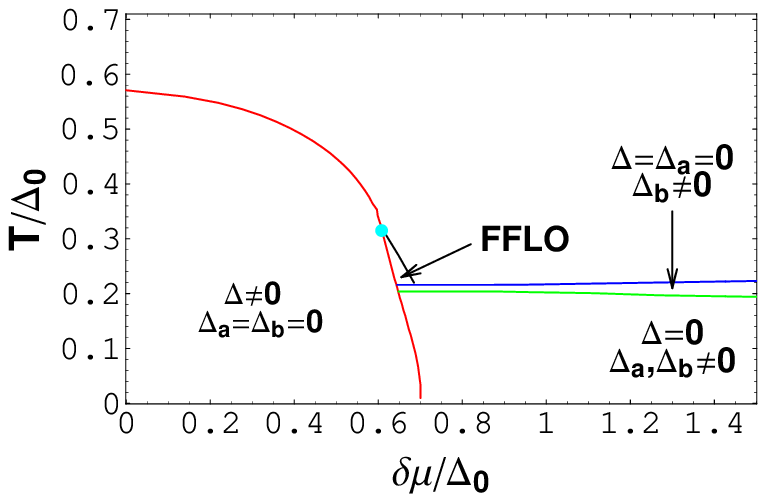}
\caption{ (color online) The phase diagram in $T-\delta\mu$ plane
at fixed $\mu=50\Delta_0$ with $\Delta_0$ being the symmetric gap
at $\delta\mu=0$. The upper panel is with only double-species
pairing, and in the bottom panel the single-species pairing is
included as well. } \label{fig1}
\end{center}
\end{figure}

The phase diagram in $T-\delta\mu$ plane is shown in
Fig.\ref{fig1}. The upper panel is in the familiar case without
single-species pairing\cite{taka}. The homogeneous BCS state can
exist at low temperature and low mismatch, and the inhomogeneous
FFLO state survives only in a narrow mismatch window. When the
single-species pairing is included as well, see the bottom panel
of Fig.\ref{fig1}, the inhomogeneous FFLO state of $a$-$b$ pairing
is eaten up by the homogeneous superfluid of $a$-$a$ and $b$-$b$
pairings at low temperature, just as we expected, and survives
only in a small triangle at high temperature. The phase transition
from the $a$-$b$ pairing superfluid to the $a$-$a$ and $b$-$b$
pairing superfluid is of first order. Note that for systems with
fixed chemical potentials there is no mixed phase of double- and
single-species pairings, and the situation is similar to a
three-component fermion system\cite{paan}.

\section{Summary}
We have investigated the phase structure of an asymmetric
two-species fermion superfluid with both double- and
single-species pairings. Since the attractive interaction for the
single-species pairing is relatively weaker, it changes the
conventional phase diagram at low temperature. In any system with
chemical potential imbalance, the double-species pairing in FFLO
state is replaced by the single-species pairing. In the region
with single-species pairing, the interesting gapless superfluid is
washed out and all fermion excitations are fully gapped. We should
note that in this paper we considered only the grand canonical
ensemble. For the case of canonical ensemble, the phase diagram
becomes much more rich\cite{huang}.
\\ \\
{\bf Acknowledgments:} The work was supported by the grants
NSFC10575058, 10425810 and 10435080.

\end{document}